\documentclass[%
reprint,
 longbibliography,
 amsmath,amssymb,
 aps,
prb,
]{revtex4-1}

\usepackage{xcolor}
\usepackage{graphicx}
\usepackage{dcolumn}
\usepackage{multirow}
\usepackage{bm}

\renewcommand{\bf}{\textbf}

\newcommand{\B}{\beta}

\begin{document}
	
	
	\title{Conformal sparse metasurfaces for wavefront manipulation}
	\author{Vladislav Popov$^1$}
	\author{Shah Nawaz Burokur$^2$}
	\email{sburokur@parisnanterre.fr}
	\author{Fabrice Boust$^{1,3}$}
	\affiliation{%
	$^1$SONDRA, CentraleSup\'elec, Universit\'e Paris-Saclay,
		F-91190, Gif-sur-Yvette, France
	}%
	\affiliation{%
	$^2$LEME, UPL, Univ Paris Nanterre,~F92410 Ville d'Avray, France
	}
	\affiliation{%
$^{3}$DEMR, ONERA, Universit\'e Paris-Saclay, F-91123, Palaiseau, France
	}

	\begin{abstract}
	The last decade was marked by the advent of the concepts of phase gradient and Huygens' metasurfaces which simple theoretical models have been leading the research on metasurfaces ever since. 
	Meanwhile, theoretical modelling of non-planar metasurfaces have appeared to be exceptionally challenging where it demands accurate analysis of the metasurface geometry. The present work addresses this challenge with a radically different theoretical approach demonstrating how numerical calculation of Green's function can be employed to design conformal metasurfaces of arbitrary geometries within the same framework and without any accommodation. 
	Against the classical concepts, the new approach permits building a metasurface of meta-atoms with electric-only response and significantly simplifying its design and fabrication.
	The theory is endorsed experimentally by testing several metasurface prototypes  for different beam-forming  functionalities: radiation with single or multiple beams in any desired directions.
	\end{abstract}

    \maketitle

	\section{Introduction}
	
	In the last two decades, metasurfaces that are thin two-dimensional metamaterials have proven themselves as a powerful tool to tailor wavefronts~\cite{Genevet2011,Grbic2013}.
	A myriad of
	metasurface-based planar devices has been proposed and validated. Nowadays there is an increasing interest in conformal metasurfaces to perfectly match  curved shapes~\cite{conformal_Nawaz2013,conformal_Gregoire2013,conformal_Liu2015,conformal_Sriram2015,conformal_Hossein2016,conformal_Faraon2016,conformal_Sun2018,Genevet2018_conformal,conformal_Denidni2019}. As expected, the design and fabrication of conformal metasurfaces  are  more  demanding compared to planar metasurfaces~\cite{conformal_Sriram2015}. 
    Adopting flexible substrates to implement metasurfaces opens a new way for integration with other elements and non-planar designs including lightweight wearable devices~\cite{conformal_Sriram2015,conformal_Faraon2016,Genevet2018_conformal}.
	In addition, for example, microwave antennas based on conformal metasurfaces do not only allow one to meet the aerodynamic specifications of aircrafts  and  satellites~\cite{conformal_Nawaz2013,conformal_Denidni2019} but also break the fundamental constraint of their planar counterparts. For instance, the aperture of a flat metasurface antenna vanishes when the beam steering angle increases~\cite{nayeri2018reflectarray}.

	The design of a conformal metasurface is generally based on geometrical optics approach~\cite{conformal_Nawaz2013,conformal_Gregoire2013,conformal_Liu2015,conformal_Sriram2015,conformal_Hossein2016,conformal_Faraon2016,conformal_Sun2018,conformal_Denidni2019}, where a proper spatial distribution of \textit{local} reflection (transmission) coefficient should be established along the reflecting (transmitting) metasurface. 
	It has been revealed that this heuristic approach has strong limitations in terms of efficiency and versatility~\cite{Asadchy2016,Alu2016}.
	On the other hand, more rigorously,  metasurfaces can be described  by means of \textit{continuous} surface impedances~\cite{Holloway2003_GSTC,Epstein:16}.
     Unfortunately, theoretical modelling of conformal metasurfaces appears to be exceptionally challenging, demanding accurate analysis of the metasurface geometry and dealing with curvilinear coordinates~\cite{Genevet2018_conformal}.

    The contribution of this work is twofold. 
    First, 
    it is shown how numerical calculation of a Green's function can be employed to design conformal \textit{sparse} metasurfaces capable of creating arbitrary field patterns for arbitrary external excitations. 
    The proposed approach does not use any complex local coordinate system matched to a particular geometry such that \textit{arbitrarily-shaped} metasurfaces can be considered without any accommodation.
    Sparse metasurfaces possess strongly non-local response~\cite{Alu2017_metagr,Epstein2017_mtg,Eleftheriades2018,Popov2019,Epstein2016_prl,Tretyakov2017_NLM,Kwon2018_NLM} and can be described in terms of neither surface impedance nor local reflection and/or transmission coefficients being not subject to fundamental efficiency limitations as their ``dense'' counterparts~\cite{Epstein2014_ieee,Asadchy2016,Alu2016,Epstein2016_prl,Epstein2016_ieee}.
    At the same time, the sparseness allows to establish a \textit{global} theoretical model and get a microscopic insight into the theoretical analysis of conformal metasurfaces.
    Secondly, it is detailed how to realize these conformal sparse metasurfaces. 
    To describe the design procedure, three semi-cylindrical sparse metasurfaces illuminated by an arbitrary complex wave configuration are experimentally demonstrated at microwave frequencies.

\section{Theory}
\label{sec:theory}
\begin{figure}[tb]
		\includegraphics[width=0.99\linewidth]{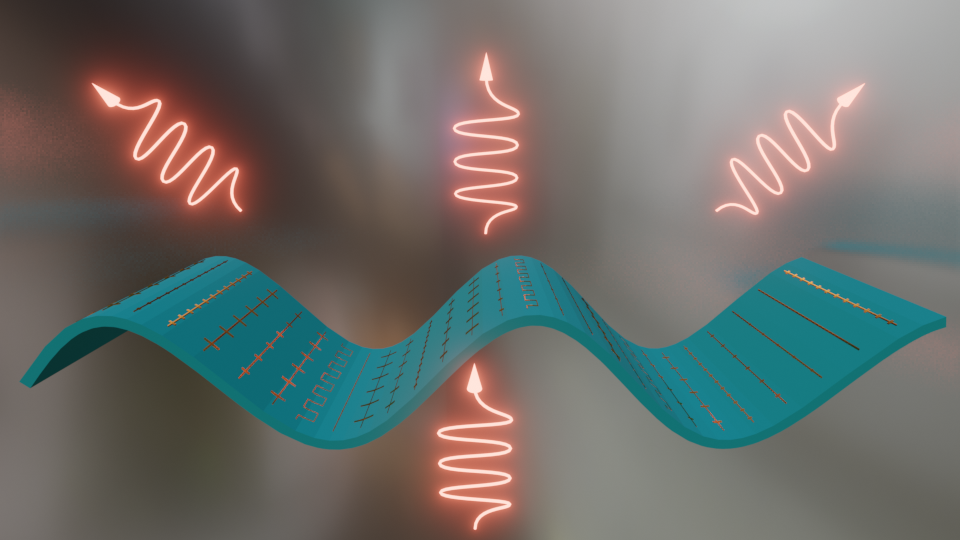}
		\caption{\label{fig:1} Illustration of an arbitrarily-shaped sparse metasurface transforming an arbitrary impinging wave into multiple beams. 
		}
	\end{figure}

	Without loss of generality, we consider the case of TE polarization and a two-dimensional (2D) geometry.
	A translation symmetry is assumed along one of the three spatial dimensions.
	We compose a sparse metasurface by a finite set of $N$ loaded wires distributed along the surface of an \textit{arbitrarily-shaped} dielectric substrate.
	The wires are oriented along the translation-invariant direction.
	An example presented in Fig.~\ref{fig:1} illustrates an array of loaded wires distributed along an arbitrarily-shaped substrate.
	Microscopically, a loaded wire represents itself as a chain of subwavelength meta-atoms.
	On the other hand, macroscopically, we model the loaded wire as uniform and having a deeply subwavelength effective radius $r_{eff}$.

	We consider external sources radiating a background wave with the electric field directed along $x$-axis and exciting polarization currents in the loaded wires.
	However, here we do not impose  any condition on the external sources unlike plane-wave excitation used for majority of metasurfaces presented in literature.
	The  polarization currents  excited in the wires can shape the field radiated by a metasurface.
	The total electric field $E_x(\bf r)$ can be split into two terms: the field $E^{(ext)}_x(\bf r)$ due to external sources alone (in the absence of wires) and the field $E^{(sct)}_x(\bf r)$ re-radiated (scattered) by the polarization currents.
	The latter is found by means of a Green's function $G_{xx}(\bf r,\bf r^\prime)$ of a system corresponding to a metasurface-based device, which means that this Green's function is not the one of free space.
	It is calculated in the presence of a curved substrate (backed with a metal sheet in the case of a reflecting configuration) and other dielectric and/or metallic parts of the system. 
	In the equation form, the Green's function is defined as
	\begin{equation}
	    \label{eq:green}
	    \left(\frac{\partial^2}{\partial y^2}+\frac{\partial^2}{\partial z^2}+k_0^2\varepsilon_r(\bf r)\right)G_{xx}(\bf r,\bf r^\prime)=i\omega\mu_0\delta(\bf r-\bf r^\prime),
	\end{equation}
	where $\bf r$ and $\bf r^\prime$ are  2D vectors in $y0z$ plane, $\mu_0$ is the permeability of free space, $k_0$ is the wavenumber of free space, $\omega$ is the angular frequency, $\varepsilon_r(\bf r)$ is the relative permittivity corresponding to dielectric and metallic parts of the metasurface-based device in the absence of the loaded wires.
	As the considered system does not have a translational symmetry in the 2D plane, a Green's function $G(\bf r,\bf r_q)$ is a function of $\bf r$ and $\bf r^\prime$ (not their difference).
	Then, the scattered electric field can be represented as the convolution
	and	since the wires are distributed only along the top face of an arbitrarily-shaped substrate,
	the integration should be performed only over this face
	\begin{equation}
	    \label{eq:conv}
	    E^{(sct)}_x(\bf r)=\int G_{xx}(\bf r,\bf r^\prime)J_x(\bf r^\prime)\textup{d}\bf r^\prime,
	\end{equation}
	where $J_x(\bf r)$ is the electric current density corresponding to polarization currents in all wires.
	Following the assumption of infinitesimally thin wires, we approximate the current density as $J_x(\bf r)=\sum_{q=1}^N I_q\delta(\bf r -\bf{r}_q)$, where $I_q$ and $\bf r_q$ is the polarization line current excited in the $q^\textup{th}$ wire and its position, respectively. 
	Taking this into account, the total electric field can be written as
	\begin{equation}\label{eq:field}
		E_x(r,\phi)=E_x^{(ext)}(r,\phi)+\sum_{q=1}^N G_{xx}(r,\phi;\textbf{r}_q)I_q,
	\end{equation}
	where $r$ and $\phi$ are polar coordinates: radius and polar angle, respectively.
	The total field $E_x(r,\phi)$ is represented by the superposition of the wave radiated by external sources $E_x^{(ext)}(r,\phi)$ and waves $G(r,\phi;\textbf{r}_q)I_q$ scattered by the wires.
	
	   	%
	\begin{figure*}[tb]
		\includegraphics[width=0.99\linewidth]{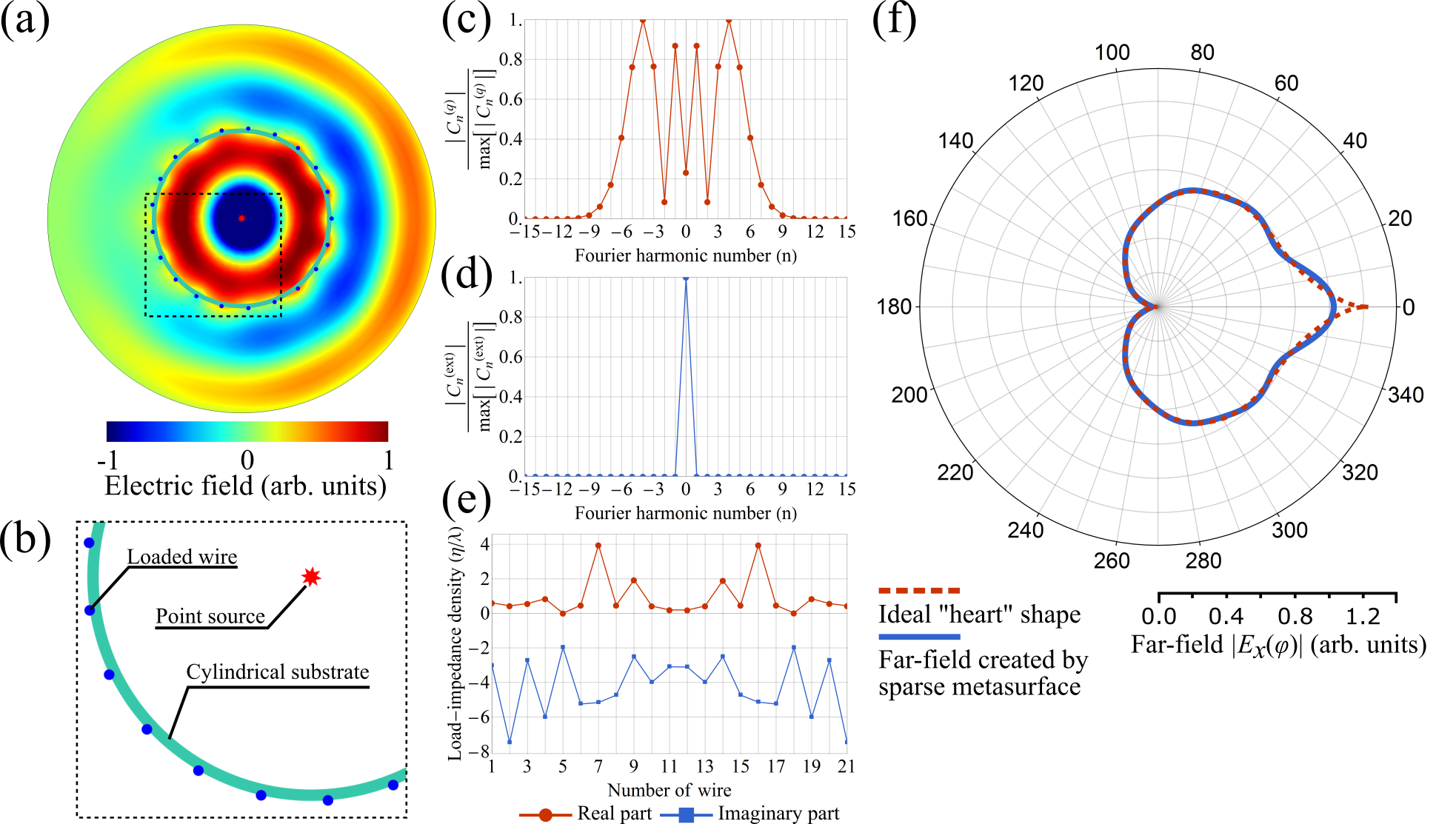}
		\caption{\label{fig:2} (a) Profile of the electric field created by a cylindrical sparse metasurface. The metasurface is represented by $21$ loaded wires uniformly distributed along a cylindrical substrate of thickness $\lambda_0/120$ and having the relative permittivity $2.2$. The metasurface is excited by a point source placed in its center. (b) Zoom of a part of the sparse metasurface. (c) $|C^{(q)}_n|/\textup{max}[|C^{(q)}_n|]$ (d)  $|C^{(ext)}_n|/\textup{max}[|C^{(ext)}_n|]$ vs. the number $n$ of Fourier harmonic. (e) Load-impedance densities required to approximate the ``heart'' shaped far-field pattern. (f) Comparison of the far-field pattern created by the sparse metasurface (solid curve) and the ideal ``heart'' shape (dashed curve).}
	\end{figure*}

	Sparse configuration of the metasurface allows one to accurately take into account the interactions between the wires via  Ohm's law
	\begin{equation}\label{eq:ohmslaw}
		Z_q I_q=E_x^{(ext)}(\textbf{r}_q)
		-\sum_{p=1}^N Z^{(m)}_{qp}I_p.
	\end{equation}
	A load-impedance density (or impedance per unit length of a wire)  $Z_q$ is a characteristic of a loaded wire and can  be engineered for instance by tuning the geometrical parameters of meta-atoms constituting a wire.
	The right-hand side of Eq.~\eqref{eq:ohmslaw} represents the total electric field at the position of the $q^\textup{th}$ wire, where 
	\begin{equation}
	    \label{eq:Zqp}
	    Z^{(m)}_{qp}=-G_{xx}(\textbf{r}_q,\textbf{r}_p)
	\end{equation}
	is the mutual-impedance density (the electric field created by the $p^\textup{th}$ wire at the position of the $q^\textup{th}$ wire).
	The separation between two neighboring wires can be arbitrary, as long as polarization currents in the wires can be approximated by a 2D delta function, which limits the transverse size of a wire.
	On the other hand, the interaction between neighboring  wires due to the macroscopic field $G_{xx}(\bf r_q,\bf r_p)$ should be much stronger than the coupling via a strongly localized microscopic near-field.
	The latter appears when geometric parts constituting two neighboring wires come very close to each other.
	As it is demonstrated further, the simple model of $\delta$ function works surprisingly well even for complex designs.
	The self-action of the $q^\textup{th}$ wire and its interaction with the environment is accounted via 
    \begin{equation}
    \label{eq:Zqq}    
    Z^{(m)}_{qq}=-\frac{1}{2\pi r_{eff}}
    \oint G_{xx}(\bf{r},\bf{r}_q)\textup{d}\textbf{r},
    \end{equation}
    where the integration is performed over the circumference of the wire of effective radius  $r_{eff}$.
	Although being very simple, Eq.~\eqref{eq:ohmslaw} has an important practical implication: it allows one to know in advance the impact of one polarization current on another and to accordingly adjust the load-impedance densities.
	Conceptually, it means that the developed approach is \textit{global}, being in strong contrast with conventional theoretical models of metasurfaces which are essentially local.
	Interaction between wires described via the matrix of mutual-impedance densities indicates that sparse metasurfaces are intrinsically strongly non-local,
	i.e. the current in the $q^\textup{th}$ wire depends on the one in the $p^\textup{th}$ wire.

    Meanwhile, by appropriately choosing $Z_q$ and the number of wires $N$, we are then able to tailor the field at will.
    Each term on the right-hand side of Eq.~\eqref{eq:field} at a given distance $r$ ($r\neq |\textbf{r}_q|$ for all $q$) can be approximated by a partial Fourier sum over the polar angle $\phi$: $E_x^{(ext)}(r,\phi)=\sum_{n=-M}^M C_n^{(ext)} e^{i n \phi}$ and 	$G_{xx}(r,\phi;\textbf{r}_q)=\sum_{n=-M}^M C_n^{(q)} e^{i n \phi}$.
	$M$ is the maximum Fourier harmonic defined as the minimum number $M$ such that 
	\begin{equation}\label{eq:def}
	|C_{\pm |M+r|}^{(ext,q)}|/\textup{max}|C_{n}^{(ext,q)}|\ll 1    
	\end{equation}
	for all $r=1,2,$... .
	The number $M$ might affect many parameters (such as geometrical parameters and material properties of a sample, the distance $r$) but the most important one is the physical aperture $\textup{max}[|\textbf{r}_q-\textbf{r}_p|]$.
	The larger the aperture the greater is $M$.
	Evidently, the total electric field $E_x(r,\phi)$ can also be represented by a partial Fourier sum $\sum_{n=-M}^M C_m e^{i n \phi}$ and the relation between the Fourier coefficients is as follows:
	\begin{equation}\label{eq:fourier}
		C_n=C_n^{(ext)}+\sum_{q=1}^N C_n^{(q)}I_q,
	\end{equation}
	where $C_n^{(ext)}$ and $C_n^{(q)}$ are known.
	When a sparse metasurface is composed of at least $N=2M+1$ loaded wires	one can establish \textit{arbitrary} azimuthal field distributions within the functional space of the $2M+1$ Fourier harmonics by adopting the Fourier coefficients $C_n$.

    Corresponding load-impedance densities can be found from Eq.~\eqref{eq:ohmslaw} after solving  Eq.~\eqref{eq:fourier} with respect to $I_q$, which in this case has a single solution~\cite{note1}.
	As a matter of fact, there is no guarantee that analytically found $Z_q$ would not require implementing active and/or lossy elements since $\Re[Z_q]\neq 0$ in a general case. 
    In order to additionally deal  with only reactive load-impedance densities $Z_q=i \Im[Z_q]$, one might need a number of wires $N\geq 2M+1$ for constructing \textit{arbitrary} radiation patterns, as recently discussed in Ref.~\cite{Popov2019}.
    Essentially, the  procedure described in this paragraph allows one to know in advance all possible configurations of the azimuthal field for a given geometry of metasurface and number of wires.
    It includes practical parameters such as beamwidth and sidelobes level.

	\begin{figure*}[tb]
		\includegraphics[width=0.99\linewidth]{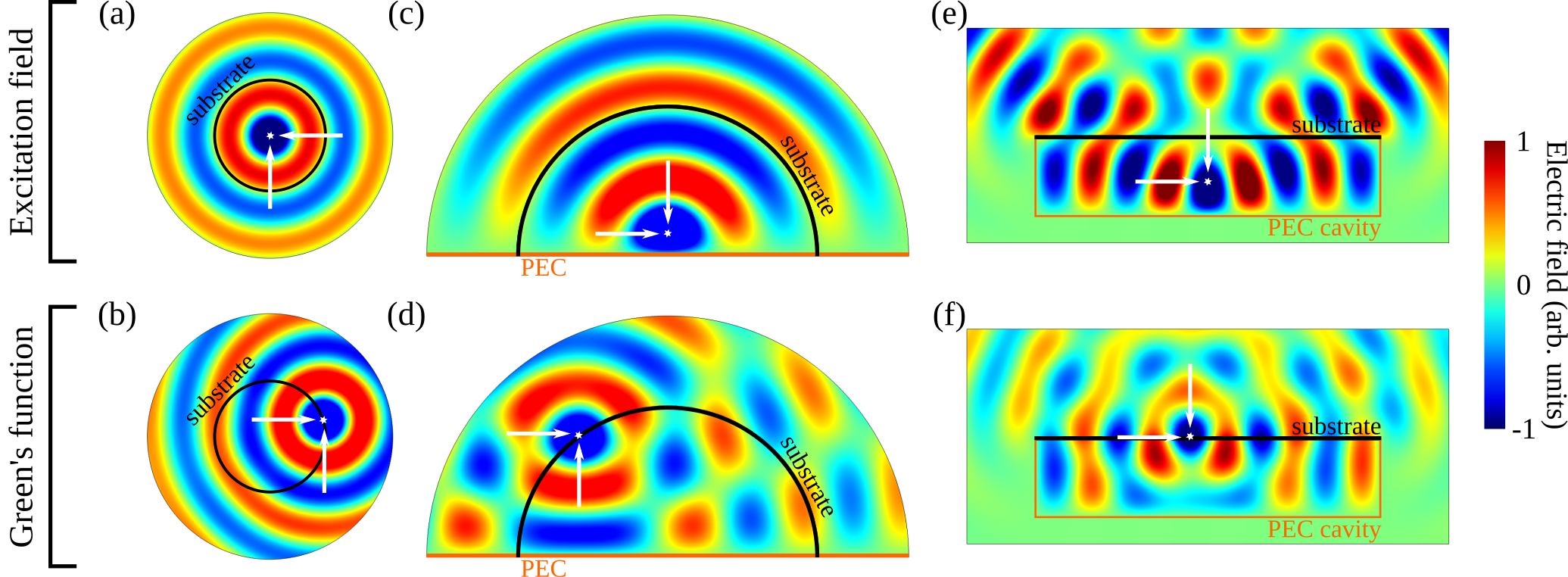}
		\caption{\label{fig:4} Simulated distributions of the electric field created by an elementary source in different configurations: (a),(b) a cylindrical dielectric substrate in free space, (c),(d) a semi-cylindrical substrate over a PEC plane, (e), (f) an open Fabry-Perot cavity cover with a planar dielectric substrate. The panels (a),(c) and (e) demonstrate distributions of the excitation field, the external (elementary) source is shown by arrows. The panels (b),(d) and (f) demonstrate function $G_{xx}(\bf r,\bf r_q)$ obtained as the electric field radiated by an elementary source located at $\bf r_q$ (shown by arrows).}
	\end{figure*}

    To sum up, after establishing the geometry of a sparse metasurface (flat, cylindrical, or any other shape), excitation type and positions of $N$ wires, we are able to calculate a Green's function $G_{xx}(r,\phi;\textbf{r}_q)$ and the background field $E_x^{(ext)}(r,\phi)$ radiated by external sources.
	With Eqs.~\eqref{eq:field} and \eqref{eq:ohmslaw}, a relation between the radiated field and load-impedance densities is established and the procedure related to Eq.~\eqref{eq:fourier} is used to find the number of wires and to determine possible functional characteristics of a sparse metasurface such as beamwidth and sidelobes level in beam-forming applications.
	A crucial point of the above analysis is the Green's function which, being defined for an arbitrary finite-size system, does not have any applicable analytical form.
	Instead, it is suggested to compute it numerically with the help of full-wave numerical simulations as detailed in Appendix~\ref{app:a}.
	A design procedure to practically implement loaded wires is described in Appendix~\ref{app:b}.
	An optimization procedure is brought to solve the inverse scattering problem assuming only \textit{reactive} load-impedance densities of wires, as detailed in Appendix~\ref{app:c}.

\section{2D simulation example}    
    \label{sec:2D}
    
    Let us consider a simple example of a cylindrical sparse metasurface for far-field manipulation.
    The radius of the metasurface is fixed to $5\lambda_0/6$ ($\lambda_0$ is the vacuum operating wavelength) and it is excited by a point source placed in the center, as shown in Figs.~\ref{fig:2}(a) and (b). 
    The source creates a cylindrical background wave, as illustrated in Fig.~\ref{fig:4}(a).
    Due to the symmetry, Green's functions $G_{xx}(r,\phi;\textbf{r}_q)$ of different wires placed on the cylindrical substrate are simply shifted with respect to each other (as function of $\phi$).
    It means that one has to analyse the Fourier decomposition of only one Green's function and of the background wave in order to find the parameter $M$.
    Figure~\ref{fig:4}(b) shows an example of the function $G_{xx}(r,\phi;\textbf{r}_q)$ computed for a particular $\bf r_q$.
    Following the definition in Eq.~\eqref{eq:def}, from Figs.~\ref{fig:2}(c) and (d), it can be seen that it is enough to have $N=2\times 10+1=21$ wires in order to be able to construct all possible  far-field radiation patterns within the Fourier space of $21$ harmonics $\textup{exp}[i n \phi]$. 
    As an illustrative example, one can reconstruct in the far-field region the shape of a ``heart'' after the required corresponding load-impedance densities are found from Eqs.~\eqref{eq:fourier} and \eqref{eq:ohmslaw}.
    The required load-impedance densities are plotted in Fig.~\ref{fig:2}(e) and correspond to passive elements ($\Re[Z_q]>0$). The real part of $Z_q$ can be engineered in a similar fashion as proposed in in Ref.~\cite{Wang2018}.
    The resulted far-field pattern is depicted in Fig.~\ref{fig:2}(f) and compared to the ideal shape of a ``heart''.  
    Figure~\ref{fig:2}(a) shows the corresponding profile of the electric field in the proximity to the metasurface.
    In order to improve the accuracy of approximating some ideal curve, one needs to increase $M$ which can be done by increasing the size of the metasurface (the radius in the considered case).
    Finally, as stated above, to best of our knowledge, there is no analytical formula for a Green's function of a system with a cylindrical substrate.
    Additional details on this example are given in  Appendix~\ref{app:a}.

\begin{figure*}[tb]
\centering
	\includegraphics[width=0.99\linewidth]{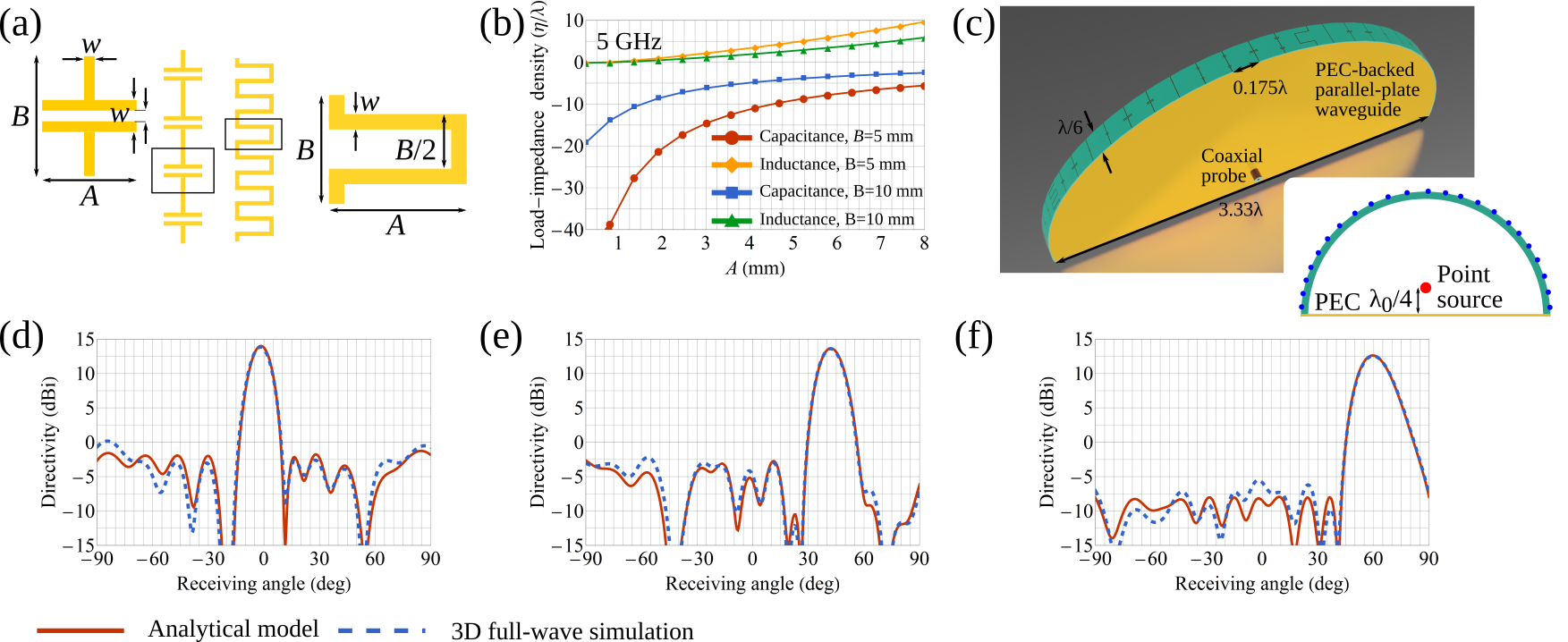}
	\caption{\label{fig:5}
	(a) Schematics of a printed capacitance (left) and inductance (right), $w=0.25$ mm ($r_{eff}=w/4$, see Ref.~\cite{Tretyakov2003}). 
	(b) Retrieved load-impedance densities (only imaginary parts are shown) of capacitively- and inductively-loaded  wires at 5 GHz. 
	(c) 3D schematic of a semi-cylindrical sparse metasurfaces of $100$ mm radius and incorporating $29$ wires. 
	The inset figure shows a 2D schematic.
	(d)--(f) Engineered radiation patterns :two-dimensional directivity vs. angle.
	Predictions of the theoretical model (solid red curves) are compared to the results of 3D full-wave simulations of three different designs of sparse metasurfaces (dashed blue curves).
	The used substrate is F4BM220 of $0.5$ mm thickness.
	Operating frequency is $5$ GHz, $\lambda_0\approx 60$ mm.
		}
\end{figure*}

\section{3D simulation examples}

In this section, we provide several examples of designs of conformal sparse metasurfaces operating in the microwave frequency range, which perform different beam-forming functionalities.
The metasurfaces demonstrated in what follows  represent a set of loaded wires uniformly distributed along the top face of a dielectric substrate. The bottom face of the substrate is metal-free. Operating in the transmission mode, the metasurfaces transform an incident wavefront from one side into a desired wavefront on the other side.
The designs are developed with the optimization-aided procedure (see Appendix~\ref{app:c}), verified by means of 3D full-wave simulations and compared to predictions of the analytical model represented by Eqs.~\eqref{eq:field} and \eqref{eq:ohmslaw}. 
The impact of using different number of wires composing a sparse metasurface is also analysed.

The design of loaded wires is performed within the local periodic approximation, developed in Ref.~\cite{Popov2019_LPA} and adopted to current examples in Appendix~\ref{app:b}.
The wires are built up from printed capacitors (left design of Fig.~\ref{fig:5}(a)) and  inductors (right design of Fig.~\ref{fig:5}(a)), which provide a wide range of accessible load-impedance densities as shown in Figs.~\ref{fig:5}(b) for $5$ GHz operation.
A particular importance of the LPA of  Ref.~\cite{Popov2019_LPA} should be emphasized: Sparse metasurfaces cannot be designed as their dense counterparts.
Indeed, the design procedure results in the load-impedance density of a loaded wire, which is its proper characteristic and depends on neither the substrate's thickness nor the inter-wire distance, see also Ref.~\cite{Popov2019IEEE}.
On the other hand, when designing conventional dense metasurfaces, one deals with such characteristics as surface impedances (or local reflection and/or transmission coefficients) which represent the integral response a unit cell and depend on the parameters of a substrate and the inter-element distance.

\subsection{Semi-cylindrical sparse metasurface}

\begin{figure*}[tb]
\centering
	\includegraphics[width=0.99\linewidth]{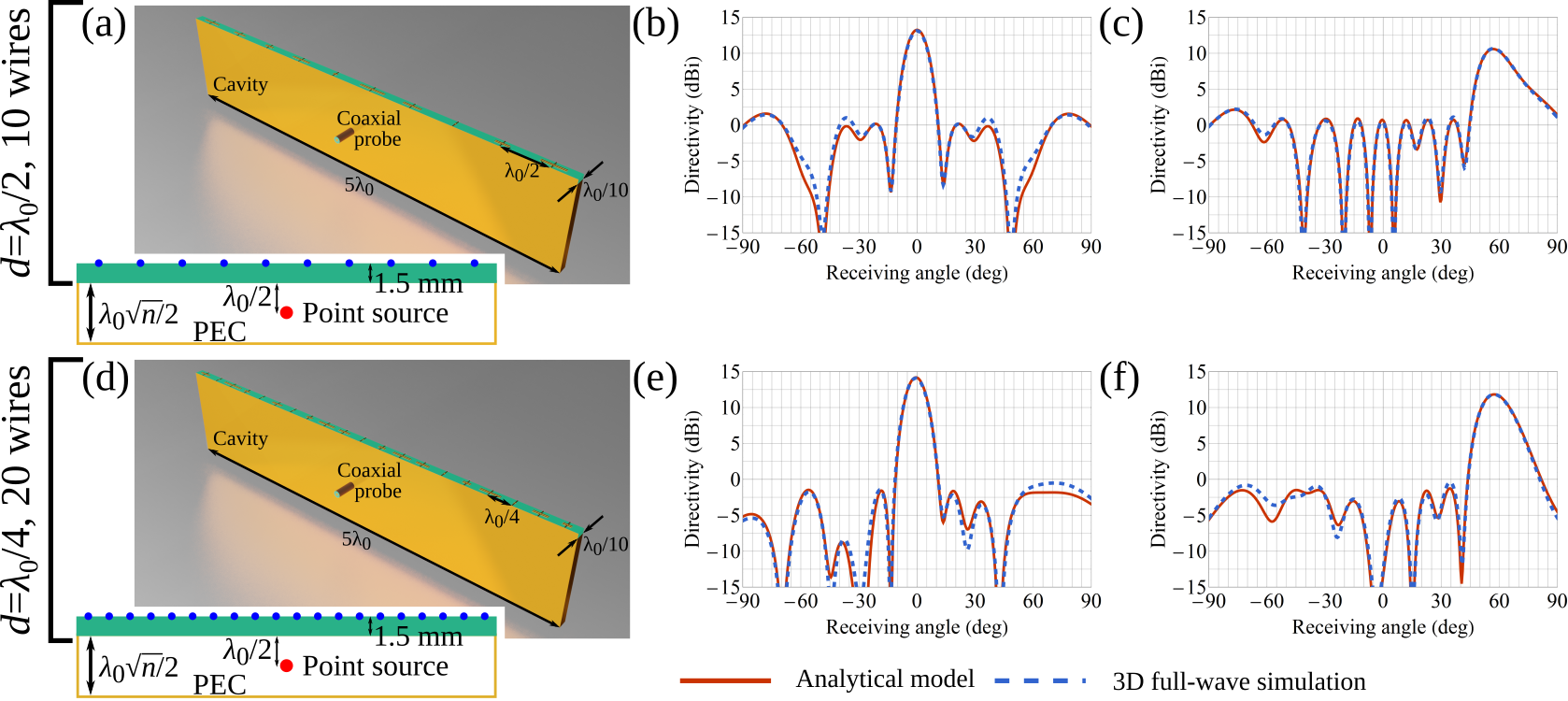}
	\caption{\label{fig:6} 
	(a), (d) 3D schematics of cavity-excited flat sparse metasurfaces having $10$ (a) and $20$ (d) wires. 
	The inset figures show 2D schematics of the corresponding configurations.
	(b),(c),(e),(f) Two-dimensional directivity vs. angle of engineered radiation patterns corresponding to analytical and 3D simulation results of four different designs of sparse metasurfaces.
	Predictions of the theoretical model (solid red curves) are compared to the results of 3D full-wave simulations of three different designs of sparse metasurfaces (dashed blue curves).
	The panels (a)--(c) and (d)--(f) correspond, respectively, to the designs incorporating $10$ wires (the inter-wire distance is $\lambda_0/2$) and $20$ wires (the inter-wire distance is $\lambda_0/4$).
	The used substrate is F4BM220 of $1.5$ mm thickness, the cavity height is $\lambda_0\sqrt{n}/2\approx 33.54$ mm, where $n$ is the cavity length ($5\lambda_0$) divided over the wavelength $\lambda_0$. 
	Operating frequency is $10$ GHz, $\lambda_0\approx 30$ mm.
	}
\end{figure*}

In the first example, we demonstrate a sparse metasurface operating at $5$ GHz (vacuum wavelength $\lambda_0\approx 60$ mm) and conformed to a semi-cylindrical shape of $1.67\lambda_0=100$ mm radius.
$N=29$ loaded wires are uniformly distributed along the top face of a $\lambda_0/120=0.5$ mm thick and $5.25\lambda_0=315$ mm long F4BM220 substrate, which serve to control $2M+1=29$ independent Fourier harmonics of the far-field as follows from Eqs.~\eqref{eq:def} and \eqref{eq:fourier}.
The metasurface is illuminated by a point source placed at $\lambda_0/4$ distance above a PEC wall joining the two ends of the semi-cylinder, as shown in the inset of Fig.~\ref{fig:5}(c).
Distribution of the electric field created by the source and an example of  function $G_{xx}(\bf r,\bf r_q)$ computed for this configuration are shown in Figs.~\ref{fig:4}(c) and (d), respectively.
In order to emulate the 2D configuration of the considered theoretical model, a narrow strip of the metasurface (of $\lambda_0/6$ width) is embedded in between two PEC plates which form a parallel-plate waveguide.
The source is represented by a coaxial probe exciting  TEM waveguide mode.
A schematics of the system is shown in Fig.~\ref{fig:5}(c).

Figures~\ref{fig:5}(d)--(f) demonstrate three different configurations of the far-field corresponding to three different designs of semi-cylindrical sparse metasurfaces.
The directivity plotted in Figs.~\ref{fig:5}(d)--(f) is the two-dimensional directivity, which is calculated as follows
\begin{equation}\label{eq:2D_directivity}
    D(\varphi)=2\pi |E_{ff}(\varphi)|^2\bigg/\int_0^{2\pi} |E_{ff}(\varphi)|^2\textup{d}\varphi,
\end{equation}
where $E_{ff}(\varphi)$ is $E_{x}(r,\varphi)$ from Eq.~\eqref{eq:field} calculated in the far-field region ($r\rightarrow\infty$).
The 3D simulation results (dashed blue curves) are compared to the corresponding far-field patterns predicted by the analytical model in Eq.~\eqref{eq:field} (solid red curves), where an almost perfect agreement is observed.
It proves a high accuracy of the established design procedure based on the analytical model represented by Eqs.~\eqref{eq:field} and \eqref{eq:ohmslaw} on one hand and the LPA on the other hand. 

Because of the very narrow $\lambda_0/6$ aperture in the transverse direction,  the radiation pattern presents a wide beam in the E-plane to produce a 3D fan-shaped beam.
However, if the distance between two PEC plates exceeds $\lambda_0/2$, higher order waveguide modes can be excited, leading to a nonuniform field distribution along the $x$-direction that was assumed to be translation invariant.
This problem can be overcome and a much narrower beam can be achieved in the E-plane by elaborating on the excitation source as discussed in the following Section.

\subsection{Cavity-excited sparse metasurface}

	\begin{figure*}[tb]
		\includegraphics[width=0.99\linewidth]{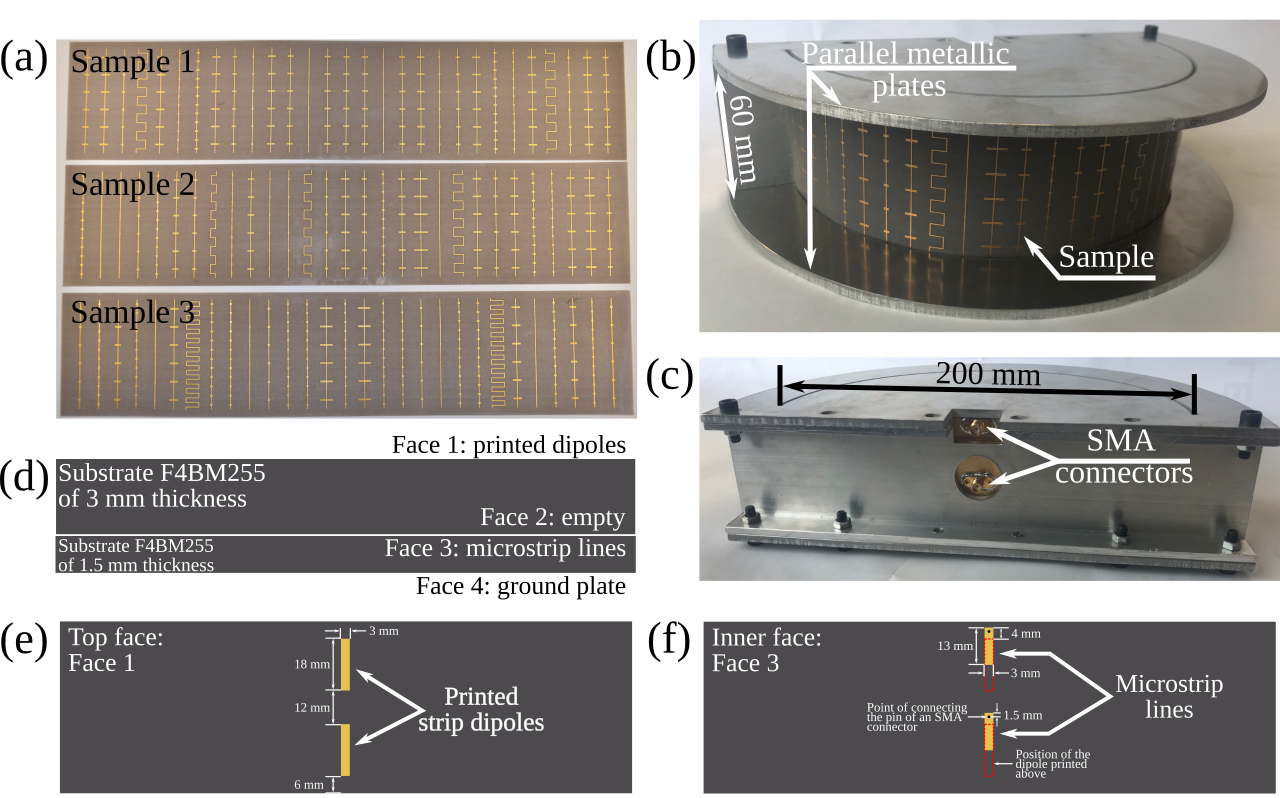}
		\caption{\label{fig:7} (a) A photography of  fabricated sparse metasurfaces on a thin flexible substrate.  (b), (c) Photographies of an assembled experimental prototype: (b) the front side and (c) the back side. 
		(d) A schematic of the side view of a PCB of the excitation source. 
		(e), (f) Schematics of the top face (e) and the inner face (f) of the PCB. 
		The schematics contain all the information necessary to design the source used to excite the samples in the experiment. 
		The design is performed following Ref.~\cite{MicrostripDipole_design}.}
	\end{figure*}
	\begin{figure*}[tb]
		\includegraphics[width=0.99\linewidth]{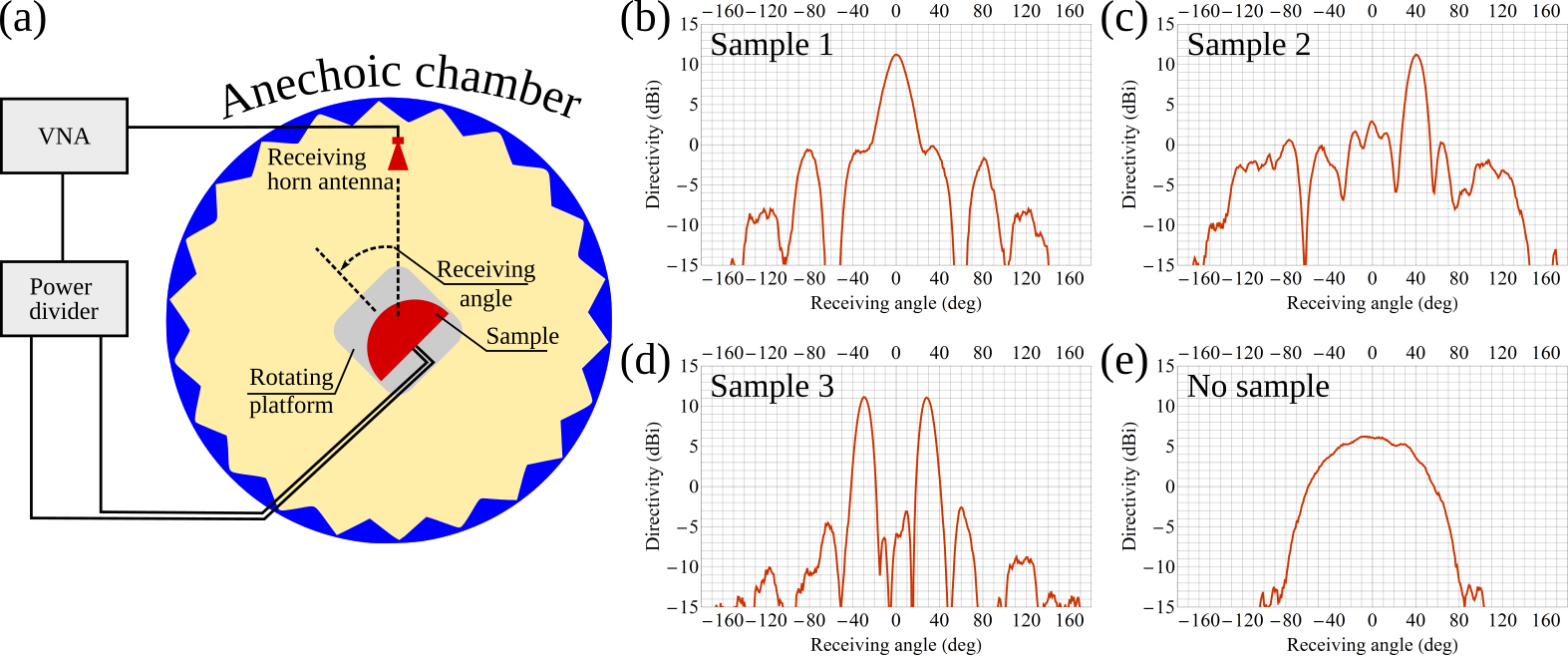}
		\caption{\label{fig:8} (a) Schematic illustration of the experimental setup used to measure the far-field patterns. (b)--(d) Measured far-field radiation patterns set up by the three fabricated samples. (e) Measured far-field radiation pattern created by the source. Operating frequency is 5 GHz.}
	\end{figure*}

In the second example, we consider planar sparse metasurfaces placed on the top of an open rectangular PEC cavity excited by a coaxial probe.
Distribution of the excitation field created by the probe inside and outside the cavity are demonstrated in Fig.~\ref{fig:4}(e).
A particular example of function $G_{xx}(\bf r, \bf r_q)$ calculated for this configuration is presented by Fig.~\ref{fig:4}(f).
Results for two configurations with $10$ and $20$ wires are compared.
For the given aperture size of $5\lambda_0$, the number $2M+1$ of independent Fourier harmonics forming the far-field equals to $42$.
While neither $10$ nor $20$ wires is enough to arbitrarily control $42$ Fourier harmonics,
it is however possible to perform efficient beam-forming.
Schematics of the configurations are shown in Figs.~\ref{fig:6}(a) and (d).
Figures~\ref{fig:6}(b),(c) and (e),(f) demonstrate engineered radiation patterns obtained by means of the developed analytical model for $10$- and $20$-element sparse metasurfaces, respectively. 
The verification is performed via 3D full-wave simulations (blue dashed curves) where an almost perfect agreement with the analytical curves (red solid) is observed.
The comparison of Figs.~\ref{fig:6}(b) and (e) and Figs.~\ref{fig:6}(c) and (f) represents an important result: two times more wires lead to a maximum of only $2$ dBi improvement of the directivity.
Therefore, one should carefully choose the number of wires as comparable performances can be achieved with a lesser effort, which can be particularly advantageous for reconfigurable designs.
Fortunately, the presented analytical model allows one to optimize the number of elements by considering a beforehand computed Green's function and without involving time-consuming 3D full-wave simulations of real designs.

For the sake of comparison with literature, we would like to mention that cavity-excited metasurfaces have been considered in works done by \textit{A. Epstein et al.}~\cite{epstein2016cavity,Epstein2016_ieee}.
In strong contrast to the single-layer design of sparse metasurfaces presented in Fig.~\ref{fig:6}, metasurfaces are designed as dense three-layer structures to emulate magnetic and/or bianisotropic response in Refs.~\cite{epstein2016cavity,Epstein2016_ieee}.

\section{Experimental examples}

\begin{figure*}[tb]
    \centering
	\includegraphics[width=0.99\linewidth]{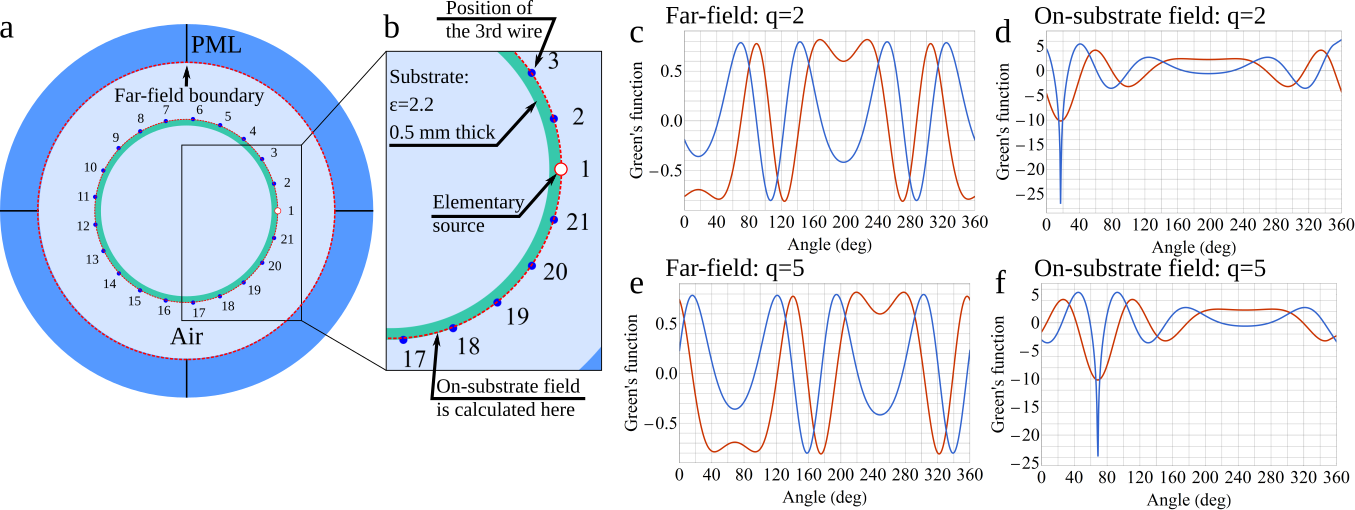}
	\caption{\label{fig:app_A} (a),(b) Two-dimensional COMSOL simulation model: a cylindrical substrate of 50 mm radius is placed in the air region surrounded by a PML layer. The model is excited by an elementary source consequently placed at the positions of the wires. The elementary source is represented by a hollow disk of radius $r_0= 0.25/4$ mm (its interior is excluded from the model) with applied at the red circle electric current density boundary condition $1/(2\pi r_0)$ A/m. (c)--(f) Green’s function calculated in the far-field region (c),(e) and on the outer face of the substrate (d),(f) regions when the elementary source is at the position of the $2^\textup{nd}$ (c),(d) and the $5^\textup{th}$ (e),(f) wires. 
	}
\end{figure*}

	To validate the approach, we designed and fabricated three sparse metasurfaces using extremely thin ($\lambda_0/240=0.25$ mm) flexible substrates F4BM220, as highlighted by the photographies in Figs.~\ref{fig:7}(a) and (b).
	The samples operating in the microwave frequency range with a central frequency fixed at $5$ GHz were fabricated by means of conventional printed circuit board (PCB) technology. 
	Each sample is composed of twenty nine wires equidistantly distributed along a $5.25\lambda_0\approx 315$ mm long substrate, making the separation between two neighbouring wires approximately equal to $\lambda_0/5$.
	The wires are built up from printed capacitors and  inductors (see Fig.~\ref{fig:5}(a)), which provide a wide range of accessible load-impedance densities as shown in Fig.~\ref{fig:5}(b).

	The ultra-thin samples are then conformed to create semi-cylindrical surfaces of $100$ mm radius.
	A photography of the assembled prototype is presented in Figs.~\ref{fig:7}(b) and (c).
	The external source exciting the samples is made of two microstrip dipole antennas printed on a metal-backed substrate~\cite{MicrostripDipole_design}, whose design is detailed in Figs.~\ref{fig:7}(d)--(f).
	The distance between the source and sparse metasurface being $100$ mm ($\approx1.7\lambda_0$), the background field pattern is neither a plane wave nor a cylindrical wave, as highlighted by Fig.~\ref{fig:8}.

	The three samples were designed to show different beam-forming examples: a single broadside beam at $0^\circ$, a steered beam at $40^\circ$ and a multibeam configuration with two beams at $\pm 30^\circ$ from broadside.
	The inverse scattering problem was solved by maximizing the power in the desired direction and minimizing the sidelobes level (with respect to the geometrical parameters $A$ and $B$ of the loaded wires) radiated by the sparse metasurface in desired directions, and the maximization procedure was implemented as particle swarm optimization~\cite{Kennedy_PSO} detailed in Appendix~\ref{app:c}.
	The different configurations were experimentally validated by radiation patterns measurements performed in an anechoic chamber. A horn antenna used a receiver is kept fixed and the assembled prototype is mounted on a rotating platform as shown in Fig.~\ref{fig:8}(a). Figures~\ref{fig:8}(b)--(d) present the experimental results obtained from the three samples.
	The level of spurious scattering (at the operating frequency) does not exceed $-12$ dB for the first sample, $-9$ dB for the second one and $-13$ dB for the third one.
	The radiation pattern of the source alone is shown in Fig.~\ref{fig:8}(e) to highlight the beam-forming capabilities of the designed metasurfaces.

\section{Concluding remarks}

    It is important to note that in the demonstrated numerical and experimental examples, the sparse metasurfaces are transmitting while possessing only electric response.
	On the other hand, conventional phase gradient approach to design transmitting metasurfaces demands implementing additionally to the electric response an effective magnetic one.
	It is necessary to suppress reflection and achieve $2\pi$-range phase response to be able to establish a required phase gradient along the metasurface~\cite{Genevet2011,conformal_Nawaz2013,conformal_Hossein2016,conformal_Faraon2016,conformal_Denidni2019}.
	A more rigorous approach is  based on engineering of electric and magnetic surface impedances~\cite{Grbic2013,Epstein2014_ieee,Epstein:16,Genevet2018_conformal} (and sometimes electro-magnetic coupling~\cite{Asadchy2015,Epstein2016_ieee,asadchy2018bianisotropic}) to manipulate wavefronts according to the equivalence theorem~\cite{Schelkunoff1936,Grbic2013}.
	Following the theory presented in this study, realizing only electric response can be sufficient for an efficient control of wavefronts that might significantly simplify the design and fabrication of wavefront manipulation devices.
	Furthermore, intrinsic strongly non-local response of sparse metasurfaces overcomes the fundamental efficiency constraint of conventional metasurfaces imposed by the conservation of normal power flow density~\cite{Alu2017_metagr,Epstein2017_mtg,Popov2019,Epstein2016_prl,Tretyakov2017_NLM,Kwon2018_NLM}. 

	To conclude, we have presented a theoretical approach that opens  the way to consequently design conformal sparse metasurfaces without appealing to a complex theory.
	Due to the versatility of the approach, one can consider different metasurface geometries and arbitrary excitation sources within the same framework.
	The theoretical analysis represented by Eqs.~\eqref{eq:def} and \eqref{eq:fourier} allows one to approach problems of superdirectivity~\cite{Schelkunoff1943} and subdiffraction focusing~\cite{Qiu2018_diff_lens}.
	Although the experiments have been performed at microwave frequencies, the theory is valid in any frequency range and may inspire research on novel applications of conformal metasurfaces.
	Particularly, sparse metasurfaces implemented on flexible substrates can be advantageous for realizing a reconfigurability mechanism based on mechanical deformations~\cite{conformal_Sriram2015}.
	It can represent a fruitful approach to create an adaptive response without complicating a design with tunable elements (which also often bring additional ohmic losses) and bias networks.
    
\section*{Acknowledgment}

The authors would like to thank Dr. Badreddine Ratni (Univ Paris Nanterre) in preparing the experimental measurement setups.

\appendix

\section{Numerical calculation of a Green's function}
\label{app:a}

After establishing the geometry of a sparse metasurface (flat, cylindrical, or any other shape), excitation type and positions of $N$ wires, one calculates a Green's function $G_{xx}(r,\varphi;\textbf{r}_q)$ and the background field $E_x^{(ext)}(r,\varphi)$ radiated by external sources.
In order to find $G_{xx}(\bf r, \bf r_q)$, we build a 2D simulation model using the commercially available finite-element-method software COMSOL Multiphysics in this work.
An elementary unit source is consequently placed at the different positions of loaded wires $\textbf{r}_q$, $q=1,2,...,N$.
More precisely, this source is a \textit{hollow} disk of radius $r_{eff}$ with an  electric surface current density used as a boundary condition and set equal to $1/(2\pi r_{eff})$ A/m.
Figures~\ref{fig:app_A}(a) and (b) demonstrate a schematics of a simulation model used to calculate the Green’s function for the example of the cylindrical spare metasurface discussed in Section~\ref{sec:2D}. 
The electric field created by the source is recorded at each position along the substrate. 
There is no need to know the Green's function over the whole 2D plane but only at certain points.
Namely, we extract the electric field at the positions of the wires to construct the matrix of mutual-impedance densities $Z_{qp}^{(m)}$ and to control azimuthal wavefront at the distance $r$.
In beam-forming applications, far-field calculations should be performed and the electric field in the far-field as a function of the azimuthal angle is recorded.
In Figs.~\ref{fig:app_A}(c)--(f), $G_{xx}(\bf r,\bf r_q)$ for the positions of the $2^\textup{nd}$ and $5^\textup{th}$ wires are shown.

\begin{figure}[tb]
    \centering
	\includegraphics[width=0.5\linewidth]{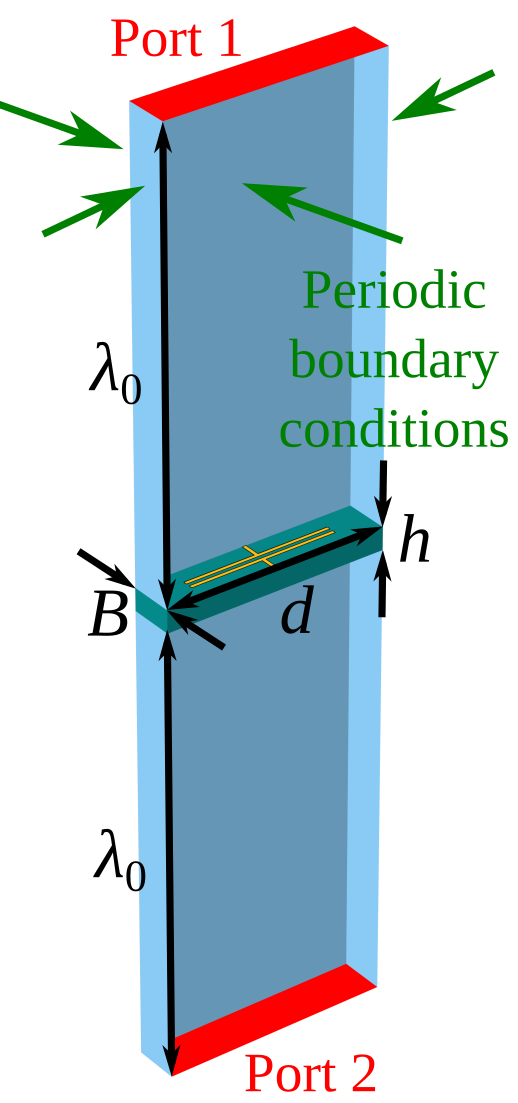}
	\caption{\label{fig:app_B}
	Principal geometry of a 3D full-wave simulation model which is used to calculate the reflection coefficient $S_{11}$ from an array of structured wires.}
\end{figure}
\begin{figure*}[tb]
\includegraphics[width=0.8\linewidth]{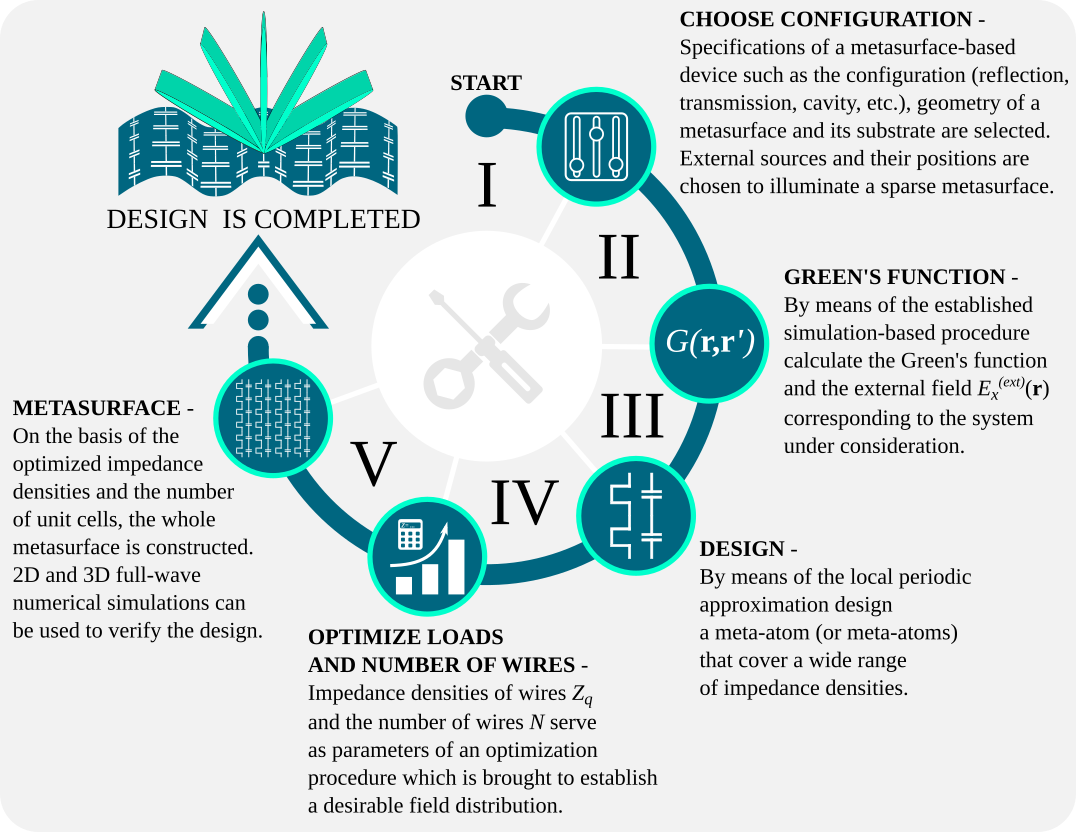}
\caption{\label{fig:app_C1} A flowchart of the optimization-aided design procedure of a sparse metasurface.}
\end{figure*}
\begin{figure*}[tb]
\includegraphics[width=0.8\linewidth]{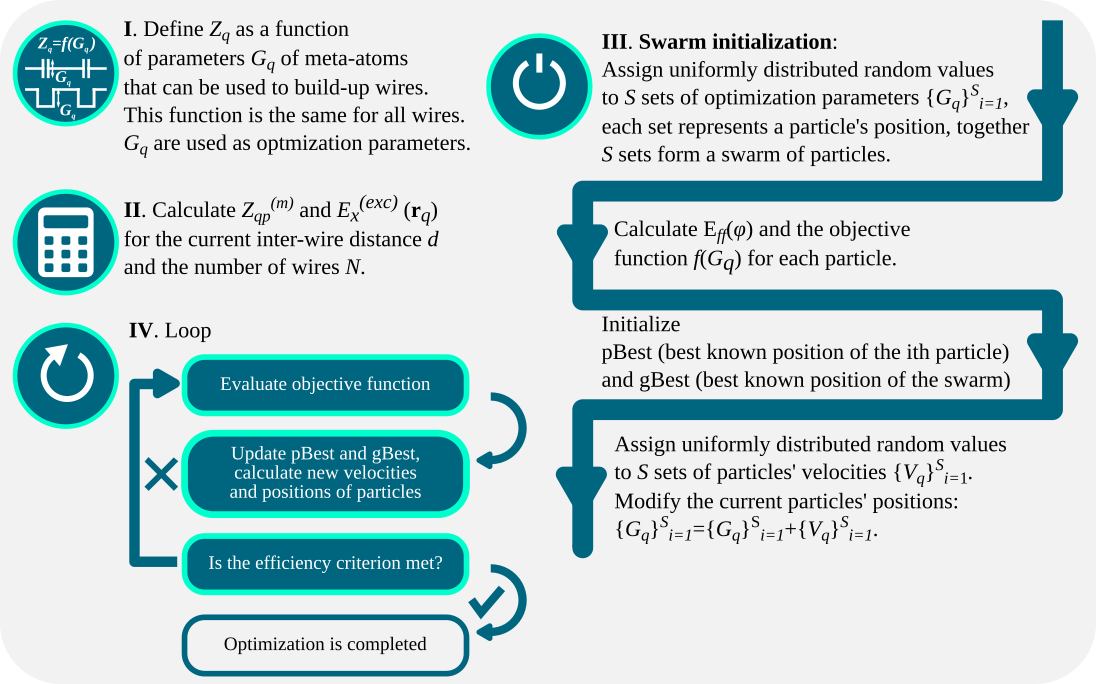}
\caption{\label{fig:app_C2} A block diagram of a particle swarm optimization algorithm employed in this work to achieve a desirable radiation pattern.}
\end{figure*}

\section{Local periodic approximation}
\label{app:b}

This Section provides details on the local periodic approximation developed in Ref.~\cite{Popov2019_LPA} for planar sparse metasurfaces in the reflection configuration.
According to the LPA, in order to find the load-impedance density, a structured wire from a nonuniform array is placed in the corresponding uniform array of period $d$ illuminated by a plane wave incident at angle $\theta$. 
It allows one to retrieve  the load-impedance density $Z_q$ from the complex amplitude of the specularly reflected TE-polarized plane wave. 
Indeed, a polarization current $I$ excited in a wire by the incident plane wave is linked to the complex amplitude $S_{11}$ of the specularly reflected wave via the following formula~\cite{Popov2019_LPA}
\begin{equation}
    \label{eq:I}
   I=-\frac{2d}{k\eta}\frac{(S_{11}e^{2i\B_0(h+\lambda_0)}-R_0^{\textup{TE}}e^{2i\B_0 h})\B_0}{(1+R_0^{\textup{TE}})e^{j\B_0 h}},
\end{equation}
where  $h$ is the thickness of the substrate, $R_m^{\textup{TE}}$ is the Fresnel's reflection coefficient from the substrate
\begin{equation}
\label{eq4:Rm_trans}
   R_m^{\textup{TE}}=
  \frac{\frac{j}{2} \left(\frac{\beta_m}{\beta_m^s} -\frac{\beta_m^s}{\beta_m}\right) \tan(\beta_m^s h)}{1+\frac{j}{2}  \left(\frac{\beta_m}{\beta_m^s} +\frac{\beta_m^s}{\beta_m}\right)\tan(\beta_m^s h)},
\end{equation}
$\beta_m=\sqrt{k_0^2-\xi_m^2}$ and $\beta_m^s=\sqrt{\varepsilon_s k_0^2-\xi_m^2}$ are,  respectively, normal components of the wavevector of the $m^\textup{th}$ Floquet mode outside and inside the substrate, $\xi_m=k_0\sin(\theta)+2\pi m/d$ is the tangential component of the wavevector.

The complex amplitude of the specularly reflected wave can be obtained from 3D full-wave simulations.
The principal geometry of a corresponding simulation model is sketched in Fig.~\ref{fig:app_B} and consists of two parts: 
a structured wire, represented by a printed capacitor on top of a dielectric substrate, and an air region.
Periodic boundary conditions are imposed on the side faces of the model.
The latter is excited by a periodic Port 1 assigned to the upper face of the air region,
which is highlighted with the red color in Fig.~\ref{fig:app_B}. 
The periodic Port 1 creates a plane wave incident at angle $\theta$ and is also used as a listening port to calculate the scattering parameter $S_{11}$.
Periodic Port 2 is set to the listening mode to accept the transmitted wave. 
The thickness of the air region above and below the substrate is set to  the operating vacuum wavelength $\lambda_0$ to eliminate higher order evanescent Floquet modes.

By modeling the polarization current excited in a structured wire as Dirac delta function $\delta(y,z)$, the interaction with the substrate and between adjacent wires can be taken into consideration analytically via the mutual-impedance density $Z_m$~\cite{Popov2019_LPA}
\begin{equation}\label{eq:Zmutual}
  Z_m=\frac{k\eta}{2}\sum_{n=1}^{+\infty} \cos[k\sin[\theta]nL]H_0^{(2)}[knd]+\frac{k\eta}{2d}\sum_{m=-\infty}^{+\infty}\frac{R_m^{TE}}{\B_m}.
\end{equation}
It allows one to retrieve the load-impedance density $Z_q$ of a wire by subtracting the electric field induced by adjacent wires and the substrate
\begin{equation}
    \label{eq:Zq}  Z_q=\frac{E_0}{I}-\frac{k_0\eta}{4}H_0^{(2)}(k_0r_{eff})-Z_m,
\end{equation}
where $E_0=(1+R_0^{TE})\exp[j\B_0h]$ represents the external electric field at the location of the central wire and induced by the incident wave and its reflection from the substrate. 
It should be noted that  the term corresponding to $Z_{qq}^{(m)}$ in Ohm's law, Eq.~\eqref{eq:ohmslaw}, accounts  for both the self-interaction and the interaction with a substrate and an environment, while
in the LPA the mutual-impedance density subtracted on the right-hand side of Eq.~\eqref{eq:Zq} corresponds only to the interaction with a substrate and other wires in the uniform array.
Therefore, Eq.~\eqref{eq:Zq} is accordingly modified by subtracting from the right-hand side the self-interaction term which, luckily, can be accounted by a simple analytical expression $-k_0\eta H_0^{(2)}(k_0r_{eff})/4$.

\section{Optimization-aided design procedure}
\label{app:c}

This Section provides a step-by-step guide to design sparse metasurfaces by means of the LPA and an optimization procedure.
The design procedure can be decomposed in five main steps, as outlined in the flowchart depicted in Fig.~\ref{fig:app_C1}.
The design procedure begins by specifying the configuration of the metasurface-based design (reflection, transmission, cavity-excited, etc.), geometry of the metasurface and its substrate.
Finally, external sources and their positions are selected to efficiently illuminate the metasurface.
Once the basic parameters are fixed, the second step is to calculate the Green's function and the external field by means of the established simulation-based procedure detailed in Appendix~\ref{app:a}.
Then, the total electric field created by a sparse metasurface is given by Eq.~\eqref{eq:field}.

The third step is independent from the previous two and requires to design a meta-atom (or meta-atoms) that will form the wires of a sparse metasurface.
To that end, the local periodic approximation is used.
A meta-atom's design should provide a wide range of achievable impedance densities of a corresponding wire when changing its geometrical parameters (or an external bias if there are embedded tunable elements).

Each desired configuration of the diffraction pattern requires different set of impedance densities which obey Ohm's law represented by Eq.~\eqref{eq:ohmslaw}.
In the fourth step of the procedure, one finds load-impedance densities of wires composing a sparse metasurface and required to establish a desired distribution of the electric field.
In general case, an arbitrary electric field $E_x(\bf r)$,
as it is stressed in Section~\ref{sec:theory}, requires impedance densities that have $\textup{Re}[Z_q]\neq 0$ and imply engineering of active and/or lossy elements. 
Including in the period additional wires and satisfying the power conservation conditions
\begin{eqnarray}\label{eq:reactive}
\textup{Re}\left[\left(E_x^{(exc)}(\bf r_q)-\sum_{p=1}^N Z_{qp}^{(m)}I_p\right)I_q^*\right]=\frac{k\eta}{4}|I_q|^2,
\end{eqnarray}
makes it sufficient to use purely reactive loads.
On the other hand, any realistic passive meta-atom possesses inevitable resistive response whether because of conduction and/or dielectric losses or embedded lossy tunable elements (such as varactor diodes, for instance).
Although the LPA allows one to account for these practical aspects of a meta-atom's design, the capacity to judiciously engineer $\textup{Re}[Z_q]$ independently from $\textup{Im}[Z_q]$ can be very limited.
Furthermore, 
the set of equations~\eqref{eq:reactive} should be modified to the following set of inequalities 
\begin{eqnarray}\label{eq3:passive}
\textup{Re}\left[\left(E_{x}^{(exc)}(\bf r_q)-\sum_{p=1}^N Z_{qp}^{(m)}I_p\right)I_q^*\right]>\frac{k\eta}{4}|I_q|^2,
\end{eqnarray}
which accounts for the resistance of wires and indicates that the power received by a wire is greater than the power it radiates.
In many situations, it is preferable to minimize the impact of $\textup{Re}[Z_q]$ resulting in increased ohmic losses.
Instead of elaborating on the analytical procedure developed in Section~\ref{sec:theory} and justifying the number of wires, the inverse scattering problem can be efficiently solved by means of a numerical optimization procedure.
It can be particularly useful for beam-forming applications  representing a subset of all possible far-field configurations when the power is maximized only in certain directions. 
Therefore, it might not be required to control all $2M+1$ of the Fourier harmonics and a lesser number of wires than $N=2M+1$ can be used for beam-forming.
As optimization parameters, the optimization procedure uses geometrical parameters of meta-atoms (or external biases) and the total number of wires.
The geometrical parameters of meta-atoms $G_q$ are related to the load-impedance density $Z_q$ of a wire by means of the LPA.
In their turn, load-impedance densities determine polarization currents and consequently the total electric field  found from Eqs.~\eqref{eq:ohmslaw} and \eqref{eq:field}.
The objective function may vary depending on the needs. The one that has been used in this work to \textit{equally} split the radiated power between an \textit{arbitrary} number of beams when minimizing the level of secondary lobes is
\begin{equation}
    f(G_q)=t^t\left(\prod_{n\in \mathbb{D}} |E_{ff}(\varphi_n)|^2\right)\bigg/
    \left(\int_0^{2\pi}|E_{ff}(\varphi)|^2\textup{d}\varphi\right)^t.
\end{equation}
Here, $\mathbb{D}$ represents a discrete set of desired beams at angles $\varphi_n$ and $t$ is the size of this set.
As a concrete optimization algorithm,  a particle swarm optimization has been employed in this work, which showed better performances than a real-valued genetic algorithm. A block diagram is displayed in Fig.~\ref{fig:app_C2}.
Once the optimal load-impedance densities are found, the fourth step of the design procedure is completed.

The fifth and the final step of the design procedure is to compose the whole metasurface out of designed meta-atoms with optimized geometrical parameters and validate the final design using 3D full-wave numerical simulations.
It is important to note that the developed optimization procedure itself does not include full-wave simulations and utilizes only numerical arrays calculated beforehand.
It makes the optimization and the development of a final design much faster than in the case when only direct optimization of a whole metasurface is performed.

\section{Geometrical parameters of experimental samples}
\label{app:d}

Table~\ref{tab:1} provides geometrical parameters of elements constituting the fabricated experimental samples.

\begin{table*}[h!]
\caption{\label{tab:1} 
Geometrical parameters of the samples. Schematics of capacitor and inductor elements is illustrated in Fig.~\ref{fig:7}(a). The parameter $w$ is the same for all elements and equals to $0.25$ mm, the thickness of the copper cladding is $35$ $\mu$m.
}
\resizebox{0.99\textwidth}{!}{%
\begin{tabular}{|c|c|c|c|c|c|c|c|c|c|}
\hline
	\multirow{2}{*}{Element number}
	 & \multicolumn{3}{c|}{Sample 1}& \multicolumn{3}{c|}{Sample 2} & \multicolumn{3}{c|}{Sample 3} \\ \cline{2-10}
	& Element type &   B (mm) & A (mm) & Element type & B (mm) & A (mm) & Element type & B (mm) & A (mm) \\ \hline
	1 & \textcolor{red}{capacitor}   & 10 & 3,31 & \textcolor{red}{capacitor} & 10 & 2,58 & \textcolor{red}{capacitor} & 10 & 1,39 \\ \hline
	2 & \textcolor{red}{capacitor}   & 10 & 4,78 & \textcolor{red}{capacitor} & 10 & 5,68 & \textcolor{red}{capacitor} & 5 & 1,04 \\ \hline
	3 & \textcolor{red}{capacitor}   & 10 & 1,58 & \textcolor{red}{capacitor} & 10 & 4,46 & \textcolor{red}{capacitor} & 10 & 3,47 \\ \hline
	4 & \textcolor{blue}{inductor}   & 10 & 5,45 & \textcolor{red}{capacitor} & 5 & 1,06 & \textcolor{red}{capacitor} & 5 & 1,30 \\ \hline
	5 & \textcolor{red}{capacitor}   & 10 & 4,81 & \textcolor{red}{capacitor} & 5 & 1,86 & \textcolor{red}{capacitor} & 10 & 0,53 \\ \hline
	6 & \textcolor{red}{capacitor}   & 5 & 0,88 & \textcolor{red}{capacitor} & 10 & 4,46 & \textcolor{red}{capacitor} & 10 & 4,72 \\ \hline
	7 & \textcolor{red}{capacitor}   & 5 & 1,99 & \textcolor{red}{capacitor} & 10 & 2,32 & \textcolor{blue}{inductor} & 5 & 7,29 \\ \hline
	8 & \textcolor{red}{capacitor}   & 10 & 4,73 & \textcolor{red}{capacitor} & 10 & 4,73 & \textcolor{red}{capacitor} & 5 & 1,53 \\ \hline
	9 & \textcolor{red}{capacitor}   & 10 & 2,41 & \textcolor{blue}{inductor} & 10 & 5,62 & \textcolor{red}{capacitor} & 5 & 1,18 \\ \hline
	10 & \textcolor{red}{capacitor}   & 10 & 1,16 & \textcolor{red}{capacitor} & 5 & 0,45 & \textcolor{red}{capacitor} & 10 & 0,56 \\ \hline
	11 & \textcolor{red}{capacitor}   & 10 & 4,31 & \textcolor{red}{capacitor} & 10 & 7,64 & \textcolor{red}{capacitor} & 10 & 1,33 \\ \hline
	12 & \textcolor{red}{capacitor}   & 10 & 3,30 & \textcolor{red}{capacitor} & 10 & 3,85 & \textcolor{red}{capacitor} & 5 & 1,07 \\ \hline
	13 & \textcolor{red}{capacitor}   & 10 & 0,60 & \textcolor{red}{capacitor} & 5 & 1,29 & \textcolor{red}{capacitor} & 5 & 1,04 \\ \hline
	14 & \textcolor{red}{capacitor}   & 10 & 3,52 & \textcolor{red}{capacitor} & 10 & 2,78 & \textcolor{red}{capacitor} & 10 & 6,89 \\ \hline
	15 & \textcolor{red}{capacitor}   & 10 & 3,42 & \textcolor{red}{capacitor} & 10 & 4,72 & \textcolor{red}{capacitor} & 5 & 0,97 \\ \hline
	16 & \textcolor{red}{capacitor}   & 10 & 3,52 & \textcolor{red}{capacitor} & 10 & 1,64 & \textcolor{red}{capacitor} & 10 & 6,89 \\ \hline
	17 & \textcolor{red}{capacitor}   & 10 & 0,60 & \textcolor{blue}{inductor} & 10 & 4,20 & \textcolor{red}{capacitor} & 5 & 1,04 \\ \hline
	18 & \textcolor{red}{capacitor}   & 10 & 3,30 & \textcolor{red}{capacitor} & 10 & 0,84 & \textcolor{red}{capacitor} & 5 & 1,07 \\ \hline
	19 & \textcolor{red}{capacitor}   & 10 & 4,31 & \textcolor{red}{capacitor} & 10 & 0,74 & \textcolor{red}{capacitor} & 10 & 1,33 \\ \hline
	20 & \textcolor{red}{capacitor}   & 10 & 1,16 & \textcolor{red}{capacitor} & 10 & 3,57 & \textcolor{red}{capacitor} & 10 & 0,56 \\ \hline
	21 & \textcolor{red}{capacitor}   & 10 & 2,41 & \textcolor{red}{capacitor} & 5 & 0,82 & \textcolor{red}{capacitor} & 5 & 1,18 \\ \hline
	22 & \textcolor{red}{capacitor}   & 10 & 4,73 & \textcolor{blue}{inductor} & 10 & 3,00 & \textcolor{red}{capacitor} & 5 & 1,53 \\ \hline
	23 & \textcolor{red}{capacitor}   & 5 & 1,99 & \textcolor{red}{capacitor} & 10 & 2,25 & \textcolor{blue}{inductor} & 5 & 7,29 \\ \hline
	24 & \textcolor{red}{capacitor}   & 5 & 0,88 & \textcolor{red}{capacitor} & 10 & 3,34 & \textcolor{red}{capacitor} & 10 & 4,72 \\ \hline
	25 & \textcolor{red}{capacitor}   & 10 & 4,81 & \textcolor{red}{capacitor} & 5 & 1,29 & \textcolor{red}{capacitor} & 10 & 0,53 \\ \hline
	26 & \textcolor{blue}{inductor}   & 10 & 5,45 & \textcolor{red}{capacitor} & 10 & 1,41 & \textcolor{red}{capacitor} & 5 & 1,30 \\ \hline
	27 & \textcolor{red}{capacitor}   & 10 & 1,58 & \textcolor{red}{capacitor} & 10 & 0,65 & \textcolor{red}{capacitor} & 10 & 3,47 \\ \hline
	28 & \textcolor{red}{capacitor}   & 10 & 4,78 & \textcolor{red}{capacitor} & 10 & 0,35 & \textcolor{red}{capacitor} & 5 & 1,04 \\ \hline
	29 & \textcolor{red}{capacitor}   & 10 & 3,31 & \textcolor{red}{capacitor} & 5 & 1,49 & \textcolor{red}{capacitor} & 10 & 1,39 \\ \hline
 \end{tabular}%
}
\end{table*}

\bibliography{bib}

\end{document}